\documentclass[sigconf]{acmart}

\usepackage{longtable}
\AtBeginDocument{%
  \providecommand\BibTeX{{%
    \normalfont B\kern-0.5em{\scshape i\kern-0.25em b}\kern-0.8em\TeX}}}

\copyrightyear{2021} 
\acmYear{2021} 
\setcopyright{acmcopyright}\acmConference[CHI '21]{CHI Conference on Human Factors in Computing Systems}{May 8--13, 2021}{Yokohama, Japan}
\acmBooktitle{CHI Conference on Human Factors in Computing Systems (CHI '21), May 8--13, 2021, Yokohama, Japan}
\acmPrice{15.00}
\acmDOI{10.1145/3411764.3445207}
\acmISBN{978-1-4503-8096-6/21/05}

\begin{document}

\title[Technology Developments in Touch Graphics 2010-20]{Technology Developments in Touch-Based Accessible Graphics: A Systematic Review of Research 2010-2020}

%% Of note is the shared affiliation of the first two authorxs, and the

\author{Matthew Butler}
\affiliation{%
  \institution{Monash University}
  \city{Melbourne}
    \country{Australia}}
\email{matthew.butler@monash.edu}

\author{Leona Holloway}
\affiliation{%
  \institution{Monash University}
  \city{Melbourne}
    \country{Australia}}
\email{leona.holloway@monash.edu}

\author{Samuel Reinders}
\affiliation{%
  \institution{Monash University}
  \city{Melbourne}
    \country{Australia}}
\email{samuel.reinders@monash.edu}

\author{Cagatay Goncu}
\affiliation{%
  \institution{Monash University}
  \city{Melbourne}
    \country{Australia}}
\email{cagatay.goncu@monash.edu}

\author{Kim Marriott}
\affiliation{%
  \institution{Monash University}
  \city{Melbourne}
    \country{Australia}}
\email{kim.marriott@monash.edu}

\renewcommand{\shortauthors}{Butler et. al.}

\begin{abstract}
This paper presents a systematic literature review of 292 publications from 97 unique venues on touch-based graphics for people who are blind or have low vision, from 2010 to mid-2020. It is the first review of its kind on touch-based accessible graphics. It is timely because it allows us to assess the impact of new technologies such as commodity 3D printing and low-cost electronics on the production and presentation of accessible graphics. As expected our review shows an increase in publications from 2014 that we can attribute to these developments. It also reveals the need to: broaden application areas, especially to the workplace; broaden end-user participation throughout the full design process; and conduct more in situ evaluation. This work is linked to an online living resource to be shared with the wider community.
\end{abstract}



\maketitle

\section{Introduction}
Since the first school for the blind was founded in Paris by Valentine Ha\"uy in 1784, raised line drawings, called tactile graphics, and 3D models have been used to convey graphical content to  people who are blind or have low vision (BLV)~\cite{eriksson1998tactile}. Tactile graphics, along with image descriptions, are now the most common method for proving accessible graphics to BLV people.
They find particular use in schools to provide BLV students with access to educational graphics such as STEM diagrams, charts and maps, and are frequently used in Orientation \& Mobility (O\&M) training.  %convey graphical content and the guidelines for their production have emerged from decades of research and practice. Tactile graphics, while arguably the most widespread and important way to convey graphical information in an accessible, explorable format, 
Tactile graphics, however, suffer several drawbacks. They require specialist equipment to produce and they are usually designed by an experienced transcriber, meaning that production is slow and costly. Thus tactiles are not well suited to, for instance, on-the-fly creation of graphics in the classroom, and because of their static, two-dimensional nature they are restricted in the types of graphical information they can directly convey. The use of tactile graphics also requires explicit training and braille literacy, limiting the number of BLV people able to effectively use them.  

Consequently, a major focus of assistive technology development has been to develop alternatives to tactile graphics that overcome some of their limitations. For instance, interactive audio labels do away with the need to understand braille and refreshable tactile displays remove the need for printing the tactile graphic. And over the past decade researchers have leveraged emerging technologies including `maker' technologies such as 3D printing. Another focus of research has been to automate or at least semi-automate the production of tactile graphics or other accessible formats. 
%And over the past decade researchers have leveraged emerging technologies including “maker” technologies such as 3D printing and low cost electronics, smartphones and apps that use crowd sourced information, computer vision, and machine learning techniques. 
%While there is a great deal of work being undertaken to explore the role of these new technologies for the provision of touch-based graphics,
But despite this activity there has been no broad critical review of the research into these new `touch-based' technologies for presenting accessible graphics to BLV people.  

This paper presents a systematic review and critique of research involving new technologies for touch-based graphics for BLV people over the period of 2010 to mid-2020. 
Systematic reviews are vital 
%to strong research communities such as CHI as 
as they allow a research field to reflect and improve on best practice \cite{brule2020review}. Our review focuses on research investigating new production or presentation methods related to touch-based graphics. By this we mean presentation methods that, like tactile graphics, rely on the reader using the position of fingers or hands to understand the spatial layout of the graphic. While other senses such as hearing can also be used, spatial understanding needs to be driven primarily by touch. 

Primary research venues for assistive and inclusive technologies were reviewed, as well as keyword searches for literature outside of these venues. In total, 352 papers were identified based on titles and abstracts. After an initial review, 59 were excluded because they did not meet the selection criteria, leaving 292 papers in the review corpus. These were analysed on the following criteria: primary research contributions; graphic content; application domain; presentation technologies; methodology, including stakeholder involvement; and availability of outcomes. 

\emph{Contributions:} This systematic review contributes to the field of accessible graphics in the following ways:
\begin{itemize}
    \item It is the first systematic review of research relating to touch-based accessible graphics. It critically reviews the work over the past decade, highlighting how the field has evolved.
    \item It summarises the results of studies comparing different touch-based presentation techniques.
    \item It identifies key omissions and concerns, revealing a number of recommendations for the assistive technologies research community. These include the need to: broaden application areas, especially to the workplace; strengthen comparative studies; conduct more authentic evaluation by increasing in situ and longitudinal studies; and broaden end-user participation in the initial design process.
    \item It provides an accessible living resource named AGRep (Accessible Graphics Repository), capturing what has been done in the field of accessible graphics technology, to be shared with the wider community and be continually updated. This repository contains all papers from this systematic review.
\end{itemize}

\section{Background}

\subsection{Provision of Touch-Based Accessible Graphics for those who are Blind or have Low Vision}
Accessibility guidelines recommend the use of tactile graphics for presenting graphics, such as maps or charts, in which spatial relationships are important \cite{BANA2010, DIAGRAM}.  Tactile graphics, however, require specialist equipment to produce and, as specialist transcribers are typically required, are  expensive and time consuming to produce. Furthermore, the reader needs sufficient tactile reading  skills, including the ability to read braille, to best understand the graphic content.

The widespread use of touch screen technologies, the arrival of low-cost maker technologies, and developments in computer vision and machine learning that can facilitate automated transcription, have opened up opportunities to take the provision of touch-based graphics into a new era. As such, new techniques for the creation of accessible graphics is a significant area of research. %Though, as yet none have been widely adopted by specialist transcribers or users. 
Key areas of research include new approaches to 3D-printed tactile models ~\cite{Carfagni2012, Grice2015, Holloway2018, Stangl2015, Teshima2010models}, haptic feedback~\cite{Darrah2013, Gay2018, Palani2017, Palani2020, Tennison2016}, robotic systems ~\cite{Guinness2019}, touch-triggered auditory feedback~\cite{Coughlan2020, Reichinger2018, Shimada2010}, and automated tactile generation \cite{wang2009instant, Chen2014, Leon2015, Taguchi2012}, such as utilising online mapping systems \cite{gotzelmann2014towards, Taylor2015, Zeng2015}.

%What is unclear from the work being conducted is its resulting impact. Indeed, transcription organisations are still heavily focused on the use of traditional techniques and mediums for delivery of touch-based graphics. This may be due to the natural lags from the emergence of a new technology or technique to when it is more widely adopted. However, it may also arise from fundamental issues in the research itself, whether in the research questions being addressed or the methodologies utilised to develop and evaluate new forms of touch-based graphics. It is therefore important to understand the nature of the work being undertaken, and reconcile this against the needs of the community to better inform research practice in this context.

%What is needed is a more holistic understanding of the work that has been conducted in advancing touch-based accessible graphics. 
Given the rapid rate at which new research is being conducted in the use of these new technologies for touch-based graphics, it is timely to review the field in order to better understand the nature of this research, and to identify best practice to better inform research in this context.

\subsection{The Value of Systematic Literature Reviews}
Systematic Literature Reviews (SLRs) are conducted across many different research fields. They are  undertaken by researchers in order to examine a research field more holistically by understanding its recent areas of focus, the gaps, and looking to the future of the field. When conducted in a rigorous manner they provide a vehicle by which a research field can reflect upon and improve  its best practice. 
%Thus, SLRs  can be significant importance by informing and inspiring the work of others.

While most research is underpinned by a review of related work, it is often very targeted and typically self-serving to the research being presented. SLRs seek to be broader in scope, synthesising “existing work in a manner that is fair and seen to be fair” \cite{kitchenham2007guidelines}, as well as remove bias \cite{mulrow1994systematic}. Not only can they provide a more complete picture of the type of work being undertaken, they can also examine a breadth of settings, method and contexts \cite{kitchenham2007guidelines, brereton2007lessons}. Thus they allow us to understand the current state of the field in a more holistic manner.

Further to this, SLRs capture a point in time and reveal what has led the research field to this point. This is important in technology-related fields, such as assistive technologies, where advances in technology and techniques have a large influence on the direction that a research field takes. 
%This is especially the case in the area of assistive technologies, where new advances in technology open up the potential to have a significant impact on the lives of those with a disability.

Finally, SLRs allow us to identify the key challenges moving forward, regarding both topic and method. They not only identify the current state of the art, but also the gaps that are yet to be addressed and provide data for rational decision making \cite{mulrow1994systematic}. They can be influential in guiding the research agenda for the years to follow. As such, “The aim of an SLR is not just to aggregate all existing evidence on a research question; it is also intended to support the development of evidence-based guidelines for practitioners.” \cite{kitchenham2009systematic}. For example, the work of Barkhuus and Rode \cite{barkhuus2007mice} presents a systematically derived set of accepted evaluation methods in the CHI community and demonstrates how they are important to specific influential research communities.

There is evidence of the benefit of systematic reviews within the domain of assistive technologies. Kelly and Smith \cite{kelly2011impact}, in their multi-decade review found that, while of benefit to the professional community, most work lacked methodological rigour. Wabiński and Moscicka \cite{wabinski2019automatic} analysed works relating to automatic tactile map production and found that very few of the working solutions reported ever made it to widespread adoption. More recently, Brulé et. al. \cite{brule2020review} sought to better understand the nature of empirical evaluation of assistive technologies for BLV users, and found that meaningful participation by BLV end users in evaluation was lacking, as well as a disproportionately low involvement of low vision participants relative to their proportion in the BLV community. Other reviews focus on a subset of work, such as the development of interactive maps \cite{Zeng2015, ducasse2018accessible}, and highlight research questions specific to that context. These works demonstrate that systematic reviews can shine a light on key issues and provoke the assistive technologies research community to reflect and improve on their practice.

\section{Methodology}
This SLR has adopted a process consistent with that recommended by Kitchenham \cite{kitchenham2007guidelines}. The process is also consistent with that reported in \cite{brule2020review}, a systematic review in a similar domain relating to BLV technology.

\subsection{Scope}
This work focused on a number of questions regarding research publications in touch-based graphics over the past decade. These are consistent with other SLRs, whereby it is the intention to capture a snapshot of current practice, identify gaps, and foreshadow the future of research in the area. Key questions are:

\begin{itemize}
    \item What are the main types of graphics and application domains studied in research exploring touch-based graphics?
    \item What are the primary presentation technologies and how have they changed over the past decade?
    \item What methodological approaches are used and how inclusive are they of the end users?
%    \item What evidence is there of sustained work and its impact on practitioners and end users?
    \item What are key gaps or challenges that researchers should consider moving forward?
\end{itemize}

This SLR aims to cover work relating to technological developments in touch-based graphics for people who are blind or have low vision (BLV). On this basis, the research team documented a number of key inclusion/exclusion examples that informed the selection of works. Notable inclusion criteria required that:
\begin{itemize}
    \item The work must relate primarily to novel touch-based graphics representations. In this context novelty refers to focus on a new production or presentation approach. A broad interpretation of `new' was adopted when considering work for inclusion;
    \item Auditory work could be considered if a touch interaction strategy was the basis;
    \item Traditional touch-based graphics, such as tactile graphics, are only included if part of a comparative study with new technologies or if exploring new production techniques. Papers solely describing the use of tactile graphics were not regarded as `novel'.
\end{itemize}

Key exclusion examples were:
\begin{itemize}
    \item Blind or low vision users are not an identified focus of the work; 
    \item Primarily focusing on braille text;
    \item Exclusive focus on user interfaces;
    \item Audio interfaces with no touch interaction.
\end{itemize}

A ten-year period was chosen for the review, rather than a longer period. This was for two primary reasons. Firstly, a large volume of relevant papers was identified in this period (over 350). It was also felt the last decade covered the emergence of topical new technologies such as 3D printing.

In order to identify relevant work, multiple search approaches were used to reduce publication bias \cite{kitchenham2007guidelines}. A systematic review of `gold standard' venues was combined with a keyword search. Gold standard venues were those that fell into one of the following categories:
\begin{itemize}
    \item Research publication venues with known interest in assistive technology, i.e.: ACM SIGACCESS Conference on Computers and Accessibility (ASSETS); ACM CHI Conference on Human Factors in Computing Systems (CHI); International Conference on Computers Helping People with Special Needs (ICCHP); PErvasive Technologies Related to Assistive Environments (PETRA); ACM Transactions on Accessible Computing (TACCESS); International Conference on Tangible, Embedded and Embodied Interaction (TEI); ACM Transactions on Computer-Human Interaction (TOCHI); and ACM Symposium on User Interface Software and Technology (UIST). These include those recognised as top-tier venues by \cite{brule2020review}.
    \item Industry journals: Journal of Visual Impairment and Blindness (JVIB); Journal of Blindness Innovation and Research (JBIR); and British Journal of Visual Impairment (BJVI).
\end{itemize}

All works in these venues were reviewed, using titles and abstracts as the primary guide. As per \cite{kitchenham2007guidelines}, a liberal and inclusive approach was taken, knowing that works could be removed at a later stage of the process. 
%To be as inclusive of work as possible, all 
All works in the venues were considered, including full papers, short papers, late-breaking works, etc.

In order to identify work published in other venues, a keyword search was also undertaken. This could capture work presented outside of venues specifically associated with assistive technologies. A Google Scholar advanced search was conducted to identify articles between 2010 and 2020, using the following keywords:
\begin{itemize}
    \item At least one of: disability; accessibility; blind; low vision;
    \item Plus at least one of: tactile; touch; haptic;
    \item Plus at least one of: graphic; diagram; map.
\end{itemize}

Rather than search specific databases e.g. ACM/IEEE, Google Scholar was used as it includes these databases. We did not use exact phrases in the search, thus the search results included synonyms. We reviewed the first 100 search results, whereby no new relevant papers were found on the next results page. After conducting the search using these two methods, 352 papers were included in the initial corpus.

\subsection{Systematic Review Process}
Based on the identified aims of the SLR, the research team created a set of analysis criteria by which each paper would be reviewed. Each criteria was discussed by the entire team to ensure agreement of its meaning was reached. Criteria were presented in a Google Form, with one form submission made for each paper.

A small cross-section of papers were reviewed by the entire team to assess the appropriateness of the analysis criteria. The research team then met to discuss the outcomes of the preliminary review, during which a revised and expanded set of analysis criteria was developed. Using a revised Google Form to capture data, all papers were reviewed against the new criteria. Results were captured for analysis in an accompanying Google Sheet. The following criteria were analysed:
contributions; graphics and their application domain; presentation technology; production methods; research methodology; evaluation methods and participation; and distribution of technology outcomes. 
%Contribution was the primary focus of the paper and its main significance to the research community. This identified if a paper was, for example, about a new presentation or production method, a comparison study, or itself a systematic review. Graphics and their intended application domain analysed the main user case of the work. For example, the graphic type may be a map, a science diagram, or be of an unspecified type. 

All 352 papers were fully reviewed. During this process each paper was also re-assessed for relevance \cite{kitchenham2007guidelines} and could be excluded from the corpus if it was deemed to be outside the scope of the literature review, as based on the initial inclusion criteria. If a reviewer believed a paper should be excluded, another reviewer would also analyse the paper. If both believed the paper should be excluded then it was formally removed from the data set. If there was disagreement, a third reviewer would make the final decision. This ensured reliability of the inclusions and exclusions \cite{kitchenham2007guidelines}. Examples of papers that were excluded at this stage include those where only widely adopted production and presentation methods with tactile diagrams were used, work where the context of application was not identified as being for BLV end users, and papers where the work only related to accessibility of user interfaces and not graphical information.

\subsection{Tools for Analysis and Sharing}
To support the authors as well as the wider touch-based graphics research community, the corpus of identified documents has been catalogued in a publicly accessible web format, based on  the interactive sentiment visualisation tool SentimentViz \cite{kucher2018state}. Like SentimentViz, our tool - AGRep (Accessible Graphics Repository) - is designed to help users to ``discover interesting patterns and facilitate data exploration'' \cite{kucher2018state}.

\begin{figure*}[h!]
    \centering
    \includegraphics[width=\textwidth]{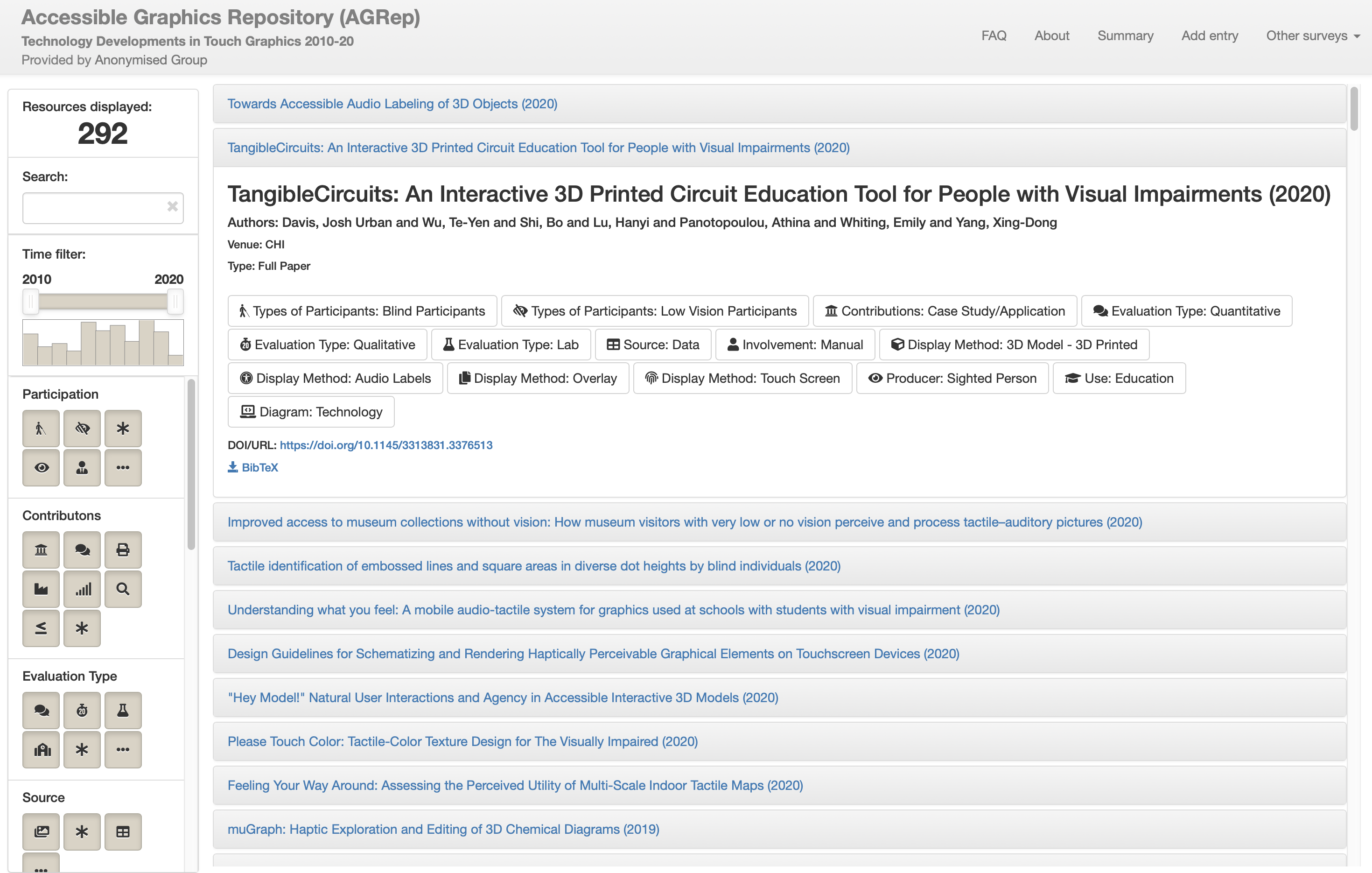}
    \Description{This image shows a screenshot of the main page of the Accessible Graphics repository website. In the main body of the page is a list of papers that are contained within the repository. One paper is expanded to show the key characteristics of the paper, as analysed in this systematic literature review. On the left of the page are filters that the user can use to view papers that meet characteristics of the users preference.}
    \caption{AGRep website}
    \label{fig:AGRep}
\end{figure*}

Using AGRep (Figure~\ref{fig:AGRep}) it is possible to explore and perform custom searches on the corpus of resources based on our categorisation markup, access persistent identifiers or URLs for each resource, and view tabulated summary data. Additionally, as AGRep is envisioned as a living resource capturing what has been done in the field of accessible graphics technology, users can update the corpus with newly published resources in the future.

\subsubsection{Accessibility Modifications}
AGRep has  been customised to improve its level of accessibility. During early stages of development the researchers held an informal session with a BLV participant to understand how the accessibility of the tool could be improved. This involved a walk-through of a test deployment of SentimentViz on the participant's personal technology. 

Based on this informal session, the following modifications were made to the source code of SentimentViz to increase its accessibility for the purpose of AGRep:
\begin{itemize}
    \item A help dialogue page added, outlining how to best use and navigate the tool;
    \item Alternate text added to all images, icons and glyph symbols;
    \item The tabbing order of page elements optimised to improve screen reader performance; 
    \item A change from grid-based layout of resources to row-based layout;
    \item Elements that update per user input marked up so that screen readers can interpret dynamically changing content.
\end{itemize}

The AGRep tool contains all papers from the corpus of this SLR, as well as more recent 2020 works that meet the inclusion criteria. The process for external parties to add new works for inclusion is given on the repository website. The AGRep tool is located at \textbf{https://agrepository.github.io}

\section{Analysis and Findings}

\subsection{Paper Counts, Venues and Authors}
On completion of the full review, the final corpus contained 292 papers from 97 unique venues. Following are the basic characteristics of the corpus. Note that the largest body of work came from ICCHP. As this conference runs every second year, data is aggregated in two-year blocks, so as not to misrepresent alternate years when ICCHP was not held.

\begin{table*}[h!]
\begin{tabular}{lcccccc}
\toprule
Year       & 2010-11 & 2012-13 & 2014-15 & 2016-17 & 2018-19 & 2020 \\
Number of Papers & 46      & 33      & 73      & 59      & 73      & 9    \\
\bottomrule
\end{tabular}\caption{\label{tab:PapersByYear}Number of Papers by Year}
\end{table*}

Table 1 shows the distribution of papers over the decade. The number of papers can be seen to have significantly risen in 2014. A possible reason for this will be discussed when we examine paper content later in this review. 
%however it appears the primary reason is the arrival of low cost “maker” technologies such as 3D printing in 2014. This led to lower research and development costs for prototyping new presentation technologies as well as interest in how these technologies could be directly used to create accessible  graphics.

\begin{table*}[!ht]
\begin{tabular}{lccccccccccc}
\toprule
Paper Count      & 1       & 2       & 3      & 4      & 5      & 6      & 7      & 10     & 35     & 44     & 61     \\
Number of Venues & 69      & 16      & 2      & 1      & 1      & 2      & 2      & 1      & 1      & 1      & 1      \\
Percentage       & 70.41\% & 16.33\% & 2.04\% & 1.02\% & 1.02\% & 2.04\% & 2.04\% & 1.02\% & 1.02\% & 1.02\% & 1.02\% \\
\bottomrule
\end{tabular}
\caption{\label{tab:PapersByVenue}Number of Papers by Venue}
\end{table*}

The corpus was drawn from 97 unique venues. As can be seen in Table 2, the vast majority of venues from which papers were chosen (70.41\%) were represented by only one paper in the corpus. Only 4 venues had on average one or more paper per year. These were: ICCHP (61 papers); ASSETS (44 papers); CHI (35 papers); and TACCESS (10 papers).

\begin{table*}[h!]
\begin{tabular}{lcccccccc}
\toprule
Paper Type & Full & Short & Poster/Demo & Thesis & Experience & Chapter & Work In Progress & Workshop \\
Count      & 171  & 59    & 37          & 7      & 6          & 4       & 4                & 4        \\
\bottomrule
\end{tabular}
\caption{\label{tab:NumPapersByType}Number of Papers by Type}
\end{table*}

The majority of papers collected were classified as ‘Full Papers’ by their venue (Table~\ref{tab:NumPapersByType}). There were also a number that were considered as ‘Short Papers’ or were an accompanying paper to a poster or demonstration. Short papers typically do not have the same expectation of rigour regarding evaluation, and include such things as preliminary studies or late-breaking work. The majority of short, poster and demo papers were in the top three venues (ICCHP, ASSETS and CHI) with a little under half being considered ‘Full Papers’ in those specific venues.

Across the 292 papers, there were 641 unique authors. All authors were counted in the metrics regardless of the author order and weightings were not applied to authorship. The large number of authors is consistent with the finding that there are a very high number of venues with only one paper (Table 2). Table 4 shows that 488 authors (76.13\%) contributed to only one paper in the corpus, i.e. have only one work relating to touch-based graphics in the last 10 years. This suggests that a significant amount of research  is a `one-off' by authors:
they explore a single idea regarding an innovation with touch-based graphics and do not further develop or evaluate it. 
In accord with this, very few authors have multiple publications. Only 32 authors (5\%) have 5 or more papers published relating to touch-based graphics over the past decade.

\begin{table*}[!ht]
\begin{tabular}{lcccccccccccc}
\toprule
Paper Count & 16     & 13     & 10     & 9      & 8      & 7      & 6      & 5      & 4      & 3      & 2       & 1       \\
Num Authors & 1      & 1      & 1      & 2      & 2      & 5      & 6      & 14     & 17     & 25     & 79      & 488     \\
Percentage  & 0.16\% & 0.16\% & 0.16\% & 0.31\% & 0.31\% & 0.78\% & 0.94\% & 2.18\% & 2.65\% & 3.90\% & 12.32\% & 76.13\% \\
\bottomrule
\end{tabular}
\caption{\label{tab:PapersByAuthor}Number of Papers by Author}
\end{table*}

%It is unclear if this consistent with other research domains. However, this 

The long tail in both venues and authors shows little evidence of `deep dives' into a particular research direction by researchers in the field. While this is somewhat speculative, it raises the concern that most authors do not have a deep knowledge of what others have done in the field and are possibly `reinventing the wheel'. It also raises the question of whether  one-off research will lead to meaningful change in BLV community practice.  On the other hand, it may be that to effect change in this area, many different approaches need to be tried to `see what sticks'. Indeed, this  was illustrated clearly in the Transforming Braille Project, initiated in 2011 to develop a low-cost refreshable braille display and bring it to market, whereby over 60 projects were examined before leading to a commercialised product \cite{TransformingBraille}. This does not, however, stop us considering the question of how work in this domain can be better shared and built upon to effect meaningful change more quickly.

\subsection{Graphics and their Application Domain}
Graphics and their application domain provide an insight into the primary use cases of the research undertaken within the corpus.

\begin{table}[!ht]
\begin{tabular}{lcc}
\toprule
Graphic Type            & \multicolumn{2}{c}{Paper Count} \\
\midrule
STEM                    & 115             &                \\
\quad Math                    &                 & 35             \\
\quad Science                 &                 & 34             \\
\quad Graphs                  &                 & 28             \\
\quad Technology              &                 & 11             \\
\quad Geometry                &                 & 5              \\
\quad Biology                 &                 & 2              \\
Maps and Plans          & 110             &                \\
General or Unspecified   & 104             &                \\
Art and Culture         & 53              &                \\
\quad Paintings               &                 & 22             \\
\quad Sculpture and Artefacts &                 & 15             \\
\quad Book Images             &                 & 8              \\
\quad Drawings                &                 & 3              \\
\quad Photographs                  &                 & 3              \\
\quad Architecture            &                 & 2              \\
Other                   & 8               &                \\
\bottomrule
\end{tabular}
\caption{\label{tab:DiagramTypes}Diagrams by Type}
\end{table}

Three categories dominated the types of graphics represented in the corpus: STEM related (115); Maps and Plans (110); and General or Unspecified (104). 
STEM related graphics encapsulated a number of other sub-types, including those relating to math, science and technology. “General or Unspecified” was a catch-all for graphics that were not explicitly defined or could be applied across domains. These included those represented by work exploring general presentation systems.
Graphics captured by ``Other'' include: Puzzles and Games (4); Screen Layout (2); and Patterns (1) and Handwriting (1).

%Their strong presence is to be expected as they are typically used within an education context. They are also a strong use case for innovation as well as being key diagrams provided by transcribers. Given the heavy reliance on tactile diagrams in Orientation and Mobility (O\&M) training, it is also unsurprising that these were a strong focus of the research collected. 

The focus on STEM related graphics and maps and plans is to be expected as it reflects that the main application areas for tactile graphics are in education and  Orientation and Mobility (O\&M) training.
Maps and plans also lend themselves to technology driven innovations. For example, they are a strong use case for 3D printing, audio labelling, and in particular automated creation, where open data is easily available. 

``Arts and Culture'' is the other main category of graphic found in the corpus. 
%Although lower than the other categories, 
This is somewhat surprising as they are not a typical use case for the provision of tactile graphics. It is exciting to see that  researchers are exploring innovation outside of the more `obvious' application areas of tactile graphics. 

We also examined if the type of graphic had varied over the past decade. We found  there was no significant change in focus. The only exception being  a small increase in graphics relating to Arts and Culture in the latter years.

\begin{table}[!ht]
\begin{tabular}{lc}
\toprule
Intended Application      & Paper Count \\
\midrule
Education         & 106         \\
O\&M              & 83          \\
Daily living      & 46          \\
Not specified     & 38          \\
Museum or gallery & 32          \\
N/A               & 7           \\
Workplace         & 5           \\
Other             & 1           \\
\bottomrule
\end{tabular}
\caption{\label{tab:DiagramUse}Diagrams by Use}
\end{table}

The intended application domain of the research publications  (Figure~\ref{tab:DiagramUse}) is consistent with the types of graphics they consider. Education and O\&M dominate, which is reflective of the %`functional' nature of the main diagrams, and 
the primary application areas for tactile diagrams. The category of ``Other'' only captured one other use case, being for research.

\subsubsection{Underrepresented Application Domains:}

 Analysis revealed a strong focus on Education and O\&M. This reflects a focus on more functional or utilitarian use, as clearly graphics in these fields provide critical information that BLV people need in their day to day lives. 
 
 Graphics that are not related to daily living, but rather quality of life across the whole life span, are much less of a focus. An exception are those relating to arts and culture. We believe an important area of future research is to consider other graphics, such as for instance, sports graphics, which are common in popular culture but currently unavailable to the BLV community.
 
An even more glaring omission is that graphics for the workplace were only considered in 5 of the papers in the corpus. This is very low and suggests significant inequities that may exist in the provision of accessible graphics in the workplace. This should be recognised as not just an opportunity for future research but one that must be addressed to support BLV people in the workplace, not just in schools or mobility training.

\subsection{Primary Contributions}

%Developments in technology over the past decade were a key motivator for undertaking the SLR. As such it is important to identify the key touch-based graphic technologies represented over the past decade and consider how that has changed during this time. It is also important to consider technologies in the context of the primary contributions being made by the works.

\begin{table}[!ht]
\begin{tabular}{lc}
\toprule
Contribution                       & Paper Count \\
\midrule
Case study or application area & 120         \\
Presentation method                & 88          \\
Production method                  & 74          \\
Interaction method                 & 69          \\
Comparative study (of presentation methods)                 & 34          \\
Other                              & 18          \\
Systematic Review                  & 7           \\
\bottomrule
\end{tabular}
\caption{\label{tab:Contributions}Paper Contributions}
\end{table}

Next we considered the area in which a paper made its primary contribution (Table~\ref{tab:Contributions}). 
A paper could make more than one contribution. 

The majority of works collected were those with a new application of existing technologies or techniques. That is not unexpected, and reflects that many works came from publication venues outside those related specifically to assistive technologies. Those works often explored the provision of touch-based graphics in a new context, for example the arts, cartography, etc. 

Behind this, three contributions shared the majority of focus of other papers. These were new presentation methods, new production methods, and new interaction methods. It is significant and positive to see that researchers are not simply focusing on new presentation methods for touch-based graphics but are also devoting equal attention to their creation and the ways in which the reader can interact with them. 

Within new production techniques, the dominant topic is the automation of accessible graphic creation, especially within the context of maps and plans. This is very important, as the timely and cost effective production of accessible graphics is a significant barrier to their more widespread use, especially in contexts outside of education and O\&M. As such it is heartening to see the strong interest by researchers but it remains an area that would benefit  from increased attention.

Other contributions that were captured by papers in the corpus included: reviews of technology (6); development or evaluation of guidelines or standards (5); exploration of design spaces and approaches (5); feasibility or usability studies (3); and understanding of tactile perception (2).

We also evaluated whether a paper was an improvement of an existing production technique or presentation technology, i.e. if it was heavily based on a preceding piece of research. While this is a more subjective measure than our other measures we believe it is still suggestive. Within the corpus, 38 studies were considered to be focused on improving existing work. This relatively low number is consistent with the `long tails' of both authors and venues as presented in section 4.1, and again should be an area of reflection for researchers.

\subsubsection{Differing Contribution Focus of  Venue:}

We discovered that different venues appear to implicitly have a slightly different focus regarding contribution. Table 8 shows the contributions of the papers in each of the 4 primary publications venues. ICCHP was more heavily skewed toward new production methods and case study areas, while ASSETS and CHI explored presentation and interaction methods more strongly. This may be consistent with the nature of ICCHP having (in general) greater involvement by practitioners. Notably, ASSETS and ICCHP had a larger number of papers that were improvement of existing techniques and technologies than the other venues.

\begin{table*}[!ht]
\begin{tabular}{lccccccc}
\toprule
        & \begin{tabular}[c]{@{}c@{}}Case study /\\ application\end{tabular} & \begin{tabular}[c]{@{}c@{}}Presentation\\ method\end{tabular} & \begin{tabular}[c]{@{}c@{}}Production\\ method\end{tabular} & \begin{tabular}[c]{@{}c@{}}Interaction\\ method\end{tabular} & \begin{tabular}[c]{@{}c@{}}Comparative\\ study\end{tabular} & \begin{tabular}[c]{@{}c@{}}Systematic\\ Review\end{tabular} & Improvement \\
\midrule
ASSETS  & 15                                                                 & 17                                                            & 12                                                          & 17                                                           & 3                                                           & 0                                                           & 10          \\
CHI     & 20                                                                 & 13                                                            & 7                                                           & 12                                                           & 2                                                           & 1                                                           & 4           \\
ICCHP   & 23                                                                 & 15                                                            & 23                                                          & 10                                                           & 3                                                           & 0                                                           & 9           \\
TACCESS & 8                                                                  & 3                                                             & 0                                                           & 3                                                            & 1                                                           & 0                                                           & 2          \\
\bottomrule
\end{tabular}
\caption{\label{tab:ContributionsByVenue}Paper Contributions by Key Venue}
\end{table*}

\subsubsection{Comparative Studies:}

Given our focus on new presentation technologies, we were interested to summarise the findings from user studies comparing different presentation technologies. We extracted these papers from the corpus and a summary of the studies and results can be seen in Table \ref{tab:comparisons} in the Appendix. Most studies focused on maps but a few used simple charts or STEM materials as stimuli. The tasks were quite varied. Almost all were laboratory studies, with only 3 in situ evaluations \cite{Koehler2017,Papadopoulos2018, Hansgen2014}.

Overall, comparisons with more than a single study support the following conclusions:
\begin{itemize}
    \item \emph{3D printed model with/without audio labels vs tactile graphic~\cite{Brittell2018,Giraud2017_1,Gual-Orti2015,Gual2015,Gual2014,Holloway2018,Koehler2017,Ramsamy-Iranah2015}}: 3D model is preferred, better short-term recall using 3D model (but not long-term recall). Some indication that the representation of third dimension is easier to understand on a 3D model. Symbols are easier to discriminate and iconic symbols more understandable with a 3D model.
    \item \emph{Touchscreen with overlay and audio labels vs tactile graphics~\cite{Brock2015,Melfi2020}}: Touchscreen with overlay and audio labels preferred and faster.
    \item \emph{Touchscreen with audio labels with/without vibration vs tactile graphic ~\cite{Giudice2012,Hahn2019,Toennies2011,Zeng2015}}: Tactile graphic is preferred and is explored more quickly. Some indication that it is difficult to understand detailed geometry with touchscreen. 
    \item \emph{Force feedback with audio labels with/without vibration vs tactile graphic~\cite{Zhang2017,Simmonet2011}}: Faster with tactile graphic. Conflicting results on accuracy.
\end{itemize}

However, this examination  revealed that more controlled studies are required to better understand the comparative advantages  of these different presentation methods. These studies should consider a wider range of graphics and more systematically investigate how low-level tasks are supported as well as the use of these methods in situ and longitudinally. 

\subsection{Presentation Technologies}

Table 9 identifies the presentation technologies used in the corpus. These technologies can be combined.
The `base' presentation technologies which provide information about the spatial layout of the graphic are, in descending order of occurrence: tactile graphics; 3D models; touch screens; refreshable tactile displays; and force feedback devices. These base technologies were frequently combined with the other presentation technologies. In the case of 3D models, audio labels were common while touch screens also provided audio labels and sometimes tactile overlays and/or vibratory feedback. 

\begin{table*}[!ht]
\begin{tabular}{lccccc}
\toprule
                            & Total & ASSETS & CHI & ICCHP & TACCESS \\
\midrule
Tactile Graphic             & 105   & 14     & 11  & 25    & 5       \\
3D Model - 3D printed       & 102   & 19     & 11  & 12    & 3       \\
Audio Labels                & 98    & 13     & 13  & 18    & 4       \\
Touch Screen                & 56    & 12     & 7   & 7     & 2       \\
Other                       & 42    & 6      & 11  & 4     & 3       \\
Refreshable Tactile Display & 37    & 5      & 5   & 10    & 1       \\
Vibration                   & 32    & 6      & 1   & 5     & 2       \\
Tactile Overlay             & 28    & 6      & 6   & 4     & 1       \\
3D Model - Other            & 24    & 3      & 1   & 5     & 2       \\
Force Feedback Device       & 21    & 2      & 1   & 6     & 3       \\
Sonification                & 18    & 4      & 3   & 2     & 3       \\
Low Vision Visuals          & 12    & 5      & 4   & 0     & 1       \\
\bottomrule
\end{tabular}
\caption{\label{tab:TechByVenue}Technologies Overall and by Key Venue}
\end{table*}

It may be surprising that tactile graphics, a well established technology, features so prominently. This reflects that they are currently the `gold standard' for accessible graphics provision and so are a frequent benchmark for comparing new presentation technologies or are the focus of new production methods. 

%When looking at the technologies present in the works, even though the inclusion criteria for literature within the corpus ensured that new developments were the focus, and that works that simply reinforced previously adopted graphic formats were excluded, tactile graphics were still the format that had the highest representation. Tactile graphics are `raised line' diagrams that have been used to provide accessible graphics for the past decades. As such, it is to be expected that they would be represented in the corpus as either the result of new production methods or in comparative studies.

%Two other technologies also dominated, namely 3D Printed Models and Audio Labels. The use of 3D printing more broadly for the creation of bespoke objects has grown dramatically in the past 5 years as 3D printing technologies have dropped in price, increased in reliability, and become more accessible to the wider community. As such this is not unexpected, especially when considering the tactile nature of 3D printed objects. In the context of this study, Audio Labels would be used in conjunction with other touch-based techniques, including 3D models and tablets, and as such their strong representation is also understandable.

\begin{figure*}
    \centering
    \includegraphics[width=\textwidth]{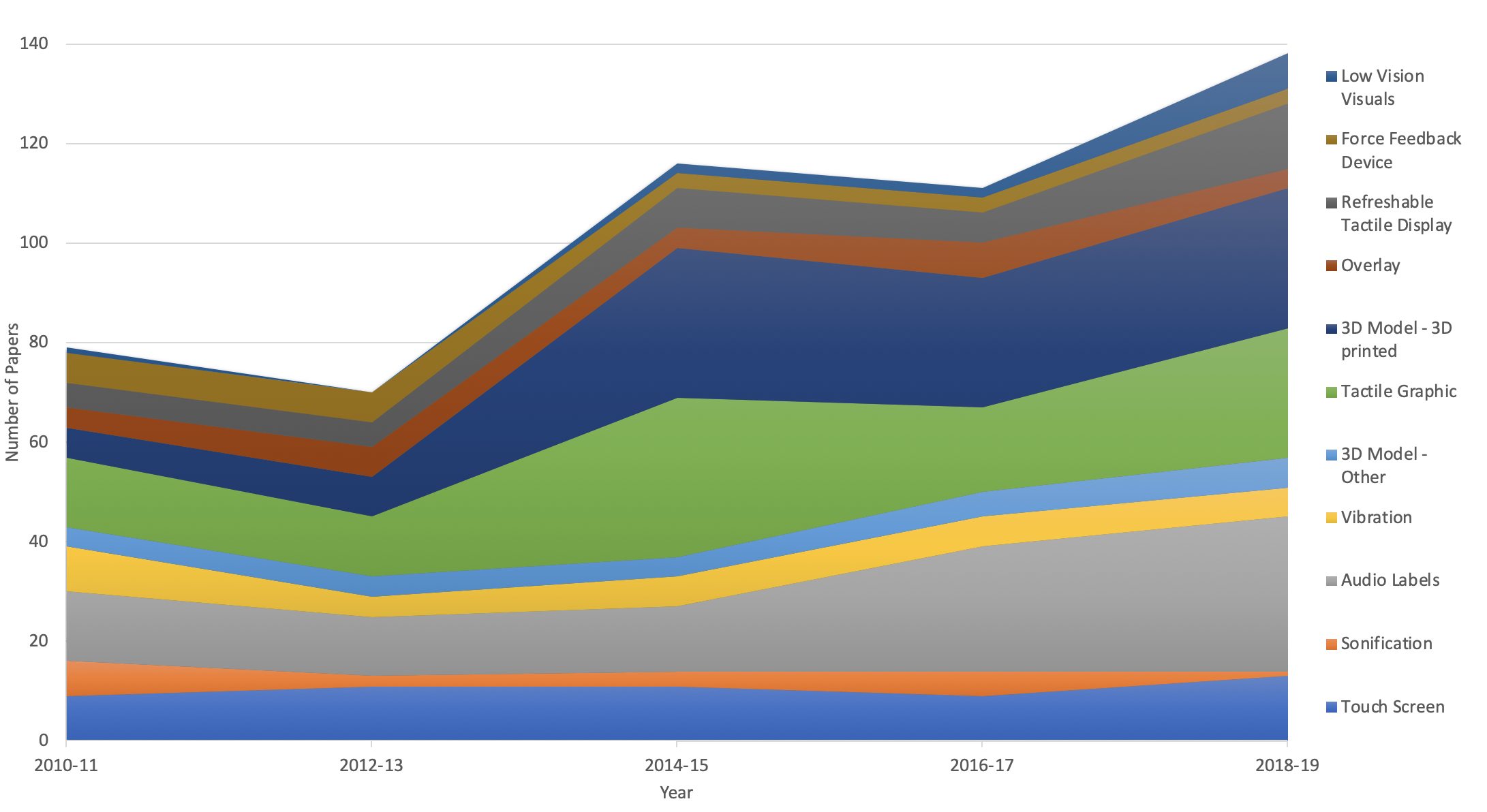}
    \Description{This image presents a stacked line chart, showing the categories of display technology, and how the number of papers about each technology have changed over time. A number of key trends can be seen, including: the number of papers increasing each year overall; a significant jump in papers on 3D printing in 2014; an increase in the use of audio labels from 2016; and a decrease in the use of force feedback devices.}
    \caption{Technology Research Over Time}
    \label{fig:TechOverTime}
\end{figure*}

\subsubsection{Emerging Technologies:}

Figure~\ref{fig:TechOverTime} shows the changes in presentation technologies over time. The most striking change is the sharp increase in the use of 3D printed models from 2014-2015. This is not unexpected, given that commodity printers arrived at this time, and over the past 5 years they have both dropped in price and been more widely adopted by  researchers and practitioners in assistive graphics provision. There has also been an associated increase in the use of audio labels, as the two technologies are frequently combined. 

We believe that the primary reason for the overall increase in publications about touch-based graphics noted earlier is the arrival of low cost `maker' technologies such as 3D printing in 2014. This led to lower research and development costs for prototyping new presentation technologies as well as interest in how these technologies could be directly used to create accessible graphics.

%Considering emerging technology, 3D printed models and Audio Labelling dominate the work. Audio labelling is reasonably consistent over the past decade (with a small rise in the latter half of the decade), which, given the ability of labelling to be provided alongside other display and interaction technologies, is unsurprising. There is, however, a clear point at which 3D printing emerged as a technology of interest, arriving dramatically in 2014-15, and continuing to have a strong presence in the work. This is not unexpected, given commodity printers arrived at this time, and over the past 5 years have both dropped in price and been more widely adopted by both researchers and touch-based graphics providers. As such, 3D printing will no doubt continue to be a strong part of work in the immediate term.

%however it appears the primary reason is the arrival of low cost “maker” technologies such as 3D printing in 2014. This led to lower research and development costs for prototyping new presentation technologies as well as interest in how these technologies could be directly used to create accessible  graphics.

\subsubsection{Technologies Out Of Favour:}

We also see that there has been diminishing attention to force feedback devices and sonification. The early part of the decade had a focus on the use of force feedback devices, such as the `Phantom Omni,' but their use dropped off almost completely by decade's end. We conjecture this is because of their expense and limited uptake by the BLV community. 
%It is unsurprising that the use of specialist hardware such as this has diminished over the past decade. These are typically expensive haptic interfaces that are not portable and have limited usability in the accessible graphics context. The embracing of 3D printing highlights that the production of bespoke physical objects is likely to continue to supersede the use of dedicated haptic interfaces, both because of the improved cost of production, quality of interaction, and engagement of experience.

The reduction in sonification research is curious. %especially as the ability to dynamically produce sonifications of data has improved in recent times. 
However, it may be that our requirement that sonification is employed in touch-based interfaces has resulted in a downward trend that is not representative of the broader research agenda. %Indeed, in this survey, the use of sonifications is explicitly tied to a touch-based interface such as an overlay, or the use of computer vision with a physical object. 
Even though this may be the case, with the availability of new ways to trigger audio labelling (such as with a moble phone, or with capacative touch sensors on 3D models), this seems a presentation technology worth pursuing.

\subsubsection{Underrepresented Technologies:}
One of the most uncommon presentation technologies is low-vision visuals, though it has increased over time. This suggests research in this area may be skewed toward people who are blind as opposed to having low-vision despite the fact that there are many more people with low vision than blindness.  This was consistent with the findings of \cite{brule2020review}. It may also reflect that the majority of people with low-vision do not wish to use touch-based graphics, however this would need to be further investigated.

%It is difficult to identify specific research areas that are underrepresented in the surveyed literature by looking solely at the technologies or research contributions. Technologies at the lower end of use in the surveyed literature include sonification and force feedback devices (as previously mentioned). Also of low representation is the use of tactile overlays. Tactile overlays could provide an excellent way to provide tactility to commodity technologies such as tablets, and as such are a research area that could benefit from greater investigation, especially when created using popular technologies such as 3D printing.

\subsubsection{Technologies by Research Contribution:}
\begin{figure*}[h!]
    \centering
    \includegraphics[width=\textwidth]{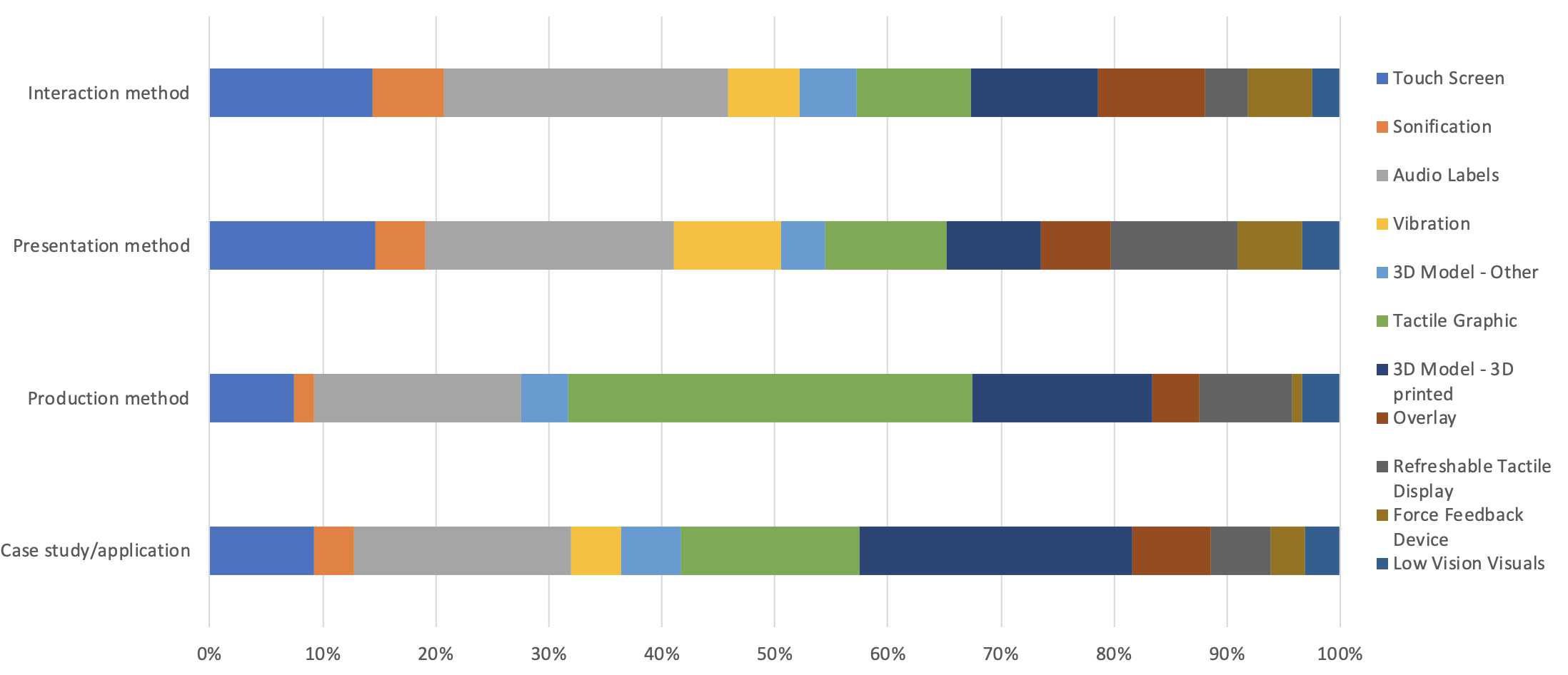}
    \Description{This image presents a stacked bar chart, showing the four main research contributions (interaction, presentation method, production method, and case study), and the proportion of papers relative to display technology. Main trends that can be seen include: significant proportion of papers with audio labels for interaction and presentation contributions; tactile graphics dominating papers on new production methods; and 3d printing being the primary technology in new case studies.}
    \caption{Technologies by Research Contribution}
    \label{fig:ContributionVersusTech}
\end{figure*}

The technologies represented in the corpus can also be considered in light of the research contributions of each work. Figure~\ref{fig:ContributionVersusTech}  shows the four main research contribution types, with the associated number of papers for each technology. Note that there can be more than one technology associated to the contribution (e.g. a case study may utilise both tactile graphics and 3D printed models). Similarly, a paper can be categorised with more than one contribution.

Examining contribution against technology reveals a number of interesting differences. When looking at case studies, 3D printing is the most common. This seems remarkable given that commodity 3D printing only emerged midway through the review period. However, it is likely that the rapid high levels of interest in 3D printing by both researchers and practitioners in accessible graphics and early positive evaluations fuelled interest in a wide variety of applications.

New production methods were dominated by tactile graphics. Within the corpus, new production methods focused heavily on automatically converting data or existing graphics  into a traditional tactile graphic. As such, this is unsurprising. However, this highlights a significant  research opportunity: How can similar automation processes  be used in the production of 3D printed models? This is especially important given the amount of work exploring the use of 3D printing for presenting touch-based graphics, and that the design of 3D models is often a barrier to their wider adoption.

We were surprised that new presentation methods were dominated by audio labels. However, this is likely due to the fact they can be used in conjunction with both touchscreens and 3D printed models. It is encouraging to see a commodity technology such as touch screens being used in the exploration of new presentation methods, as well as the more recently emerging refreshable tactile displays. 3D printed models are somewhat underrepresented in this contribution category. This is likely due to their more recent emergence; they are first being evaluated for their efficacy in different application areas before being fully explored as the basis for new presentation methods. 

%Interaction Methods were also dominated by audio labels. This is to be expected, as multi-modal interfaces are of strong interest across the entire CHI community, and the combination of tactile and auditory feedback is vital for BLV users. 

\subsection{End-User Participation}
We were particularly interested to understand the level of involvement of end users in the research. While BLV people are the primary end users, end users also include sector professionals such as transcribers or O\&M trainers.

\begin{table*}[!ht]
\begin{tabular}{lllccc}
\toprule
Participants                                                   &               & Stage     & Number of Studies & \% of Total Papers & Total Participants \\
\midrule
BLV                                                            & Blind         & Design    & 36                & 12.29\%              & 344                \\
                                                               &               & Prototype & 48                & 16.38\%              & 234                \\
                                                               &               & Eval      & 150               & 51.19\%              & 1516               \\
                                                               & Low Vision    & Design    & 13                & 4.44\%               & 85                 \\
                                                               &               & Prototype & 15                & 5.12\%               & 73                 \\
                                                               &               & Eval      & 60                & 20.48\%              & 296                \\
\midrule
Professionals                                                   & Professionals & Design    & 20                & 6.83\%               & 416                \\
                                                               &               & Prototype & 10                & 3.41\%               & 107                \\
                                                               &               & Eval      & 16                & 5.46\%               & 133                \\
\midrule
Sighted                                                        & Sighted       & Design    & 5                 & 1.71\%               & 135                \\
                                                               &               & Prototype & 11                & 3.75\%               & 98                 \\
                                                               &               & Eval      & 17                & 5.80\%               & 288                \\
                                                               & Blind Folded  & Design    & N/A               &                    & N/A                \\
                                                               &               & Prototype & 8                 & 2.73\%               & 61                 \\
                                                               &               & Eval      & 26                & 8.87\%               & 432                \\
\bottomrule
\end{tabular}
\caption{\label{tab:Participants}Participants}
\end{table*}

Table~\ref{tab:Participants} shows the involvement of participants in each research study, across the design, prototype and evaluation stages of each study. The participants are divided into three categories: BLV, Professional and Sighted. Note that participants might take part across multiple stages within a single study. Similarly, members from multiple participant groups may be used within a single stage within a study (e.g. both blind and low vision participants within an evaluation). 

Importantly, BLV participants are by far the most represented group, with the strongest representation by blind participants in the evaluation stage. One thing that is noticeable is that BLV participants, however, are much less likely to be involved in the design or prototyping stages than the final evaluation. This is potentially a serious problem as it means end user feedback is only considered late in the development. The following section explores BLV representation further.

\subsubsection{BLV and Professional Involvement:} 
%It is of greater use to look at the number of studies where there is representation by at least one group from within BLV community. 

\begin{table*}[h!]
\begin{tabular}{lccc}
\toprule
Participant Type & Number of Papers & \% of Total Papers & Total Participants \\
\midrule
BLV              & 184               & 62.80\%              & 3204               \\
BLV + Professional     & 196               & 66.89\%              & 3227      \\
\bottomrule
\end{tabular}
\caption{\label{tab:BLVParticipation}Participation by End Users (BLV and Sector Professionals)}
\end{table*}

Participation by BLV people across the corpus was in 62.80\% of the studies, which may be considered low for work that is intended to be for the benefit of a particular community. This increases marginally if sector professionals are included, to 66.89\% (Figure~\ref{tab:BLVParticipation}).

However, this must be considered in light of the paper type (Figure~\ref{tab:ParticipantsByVenue}). Short papers, posters and demonstrations had lower representation. This is primarily a constraint of either the space afforded the work, or that it is often the initial stages of new work. Full papers present a better insight into end user involvement, and as such are the focus of discussion. When only considering full papers, the involvement of BLV people in the research increases to 73.26\%. While this is strong, there is still considerable scope for improvement.

\begin{table*}[h!]
\begin{tabular}{lcccccccc}
\toprule
        & \begin{tabular}[c]{@{}c@{}}Num\\ Papers\end{tabular} & \begin{tabular}[c]{@{}c@{}}BLV\\ Involvement\end{tabular} & Percentage & \begin{tabular}[c]{@{}c@{}}Num Full\\ Papers\end{tabular} & \begin{tabular}[c]{@{}c@{}}Full Papers\\ BLV\end{tabular} & Percentage & \begin{tabular}[c]{@{}c@{}}Full BLV\\+ Prof. \end{tabular} & Percentage\\
\midrule
ASSETS  & 44                                                         & 27                                                        & 61.36\%    & 16                                                              & 14                                                        & 87.50\%    & 16                                                            & 100.00\%    \\
CHI     & 35                                                         & 28                                                        & 80.00\%    & 18                                                              & 17                                                        & 94.44\%    & 18                                                            & 100.00\%    \\
ICCHP   & 61                                                         & 40                                                        & 65.57\%    & 39                                                              & 27                                                        & 69.23\%    & 27                                                            & 69.23\%    \\
TACCESS & 10                                                         & 9                                                         & 90.00\%    & 10                                                              & 9                                                         & 90.00\%    & 10                                                            & 100.00\%   \\
Other   & 143                                                        & 80                                                        & 55.94\%    & 89                                                              & 59                                                        & 73.75\%    & 60                                                            & 75.00\%   \\
\bottomrule
\end{tabular}
\caption{\label{tab:ParticipantsByVenue}Participation by Venue}
\end{table*}

%In venues not normally associated with work relating to assistive technologies or accessibility, involvement was relatively strong at 73.75\% of papers. This is a positive finding, given that focus of the work was typically about placing the production or presentation method into a new context, and as such thorough evaluation was not always a focus or expectation. 

It could be expected that the primary venues associated with assistive technologies would have a higher involvement by the BLV community. When considering full papers only, this is mostly the case, with ASSETS, CHI and TACCESS all showing levels of engagement by BLV pople in over 85\% of papers. Indeed, when sector professionals  community are considered as well, all three venues have 100\% involvement. ICCHP, however, is an exception. BLV involvement is still only in the order of 69\% and there is no involvement with professionals within the corpus. Involvement by end users in the other venues, including venues not normally associated with assistive technologies or accessibility, is lower than the top 3 venues. 
%Given the breadth of methodological approach and possibly access to the BLV community, this is to be expected. 
Of note is that only 1 publication out of 143 papers in other venues had the involvement of a sector professional. 

One concern that emerges is that studies are skewed toward blind participants as opposed to participants with low vision. This may reflect the lack of focus on low-vision visuals that we previously identified. It is, however, consistent with the findings of \cite{brule2020review, szpiro2016people} and may reflect that researchers believe touch-based graphics are of greater importance to those who are blind. It may also be that new kinds of touch-based graphics such as touchscreens with audio labels or 3D printed models will be of benefit to people with low vision, however a focus on blind participants more stringently evaluates the technology.  
%If sothis imbalance in participants should be addressed.

%ASSETS, CHI and TACCESS are to be commended for their inclusive expectations and practice. This review should present a call to action for other venues to ensure that when research is being conducted for a specific end user group that the end users are engaged meaningfully in the development and evaluation of the work.

%\subsubsection{Creation of Diagrams:}

% Of the 239 papers that had human involvement in the creation of the touch-based graphics, only 57 (24\%) had the involvement of someone who was BLV. The involvement of BLV people as part of the diagram creation process is an important area for future work. It is not only a compelling area of research, but it is also important for supporting both analytic and artistic development.

\subsubsection{Participant Numbers:}

The numbers of participants in each study is also variable. In studies presented in full papers, there was a median of 10 BLV participants. While this may appear to be a relatively low number of participants, it is understandable given the low incidence of blindness and the difficulty that can arise with recruiting participants. With small sample sizes it is difficult to find statistical significance, so this low number may account for the greater number of qualitative studies (161) that were conducted in the corpus in comparison to quantitative studies (111). However, this is not necessarily a problem, as qualitative studies can provide rich data and important insights. Indeed \cite{brule2020review} warns against rejecting papers based on participant numbers in this field.

\subsubsection{Lack of In-Situ and Longitudinal Studies:}

One noticeable limitation of many of the studies undertaken was that they were undertaken in controlled environments, such as a laboratory (148 studies) rather than in the field or in situ (47 studies). While research that involves prototype bespoke technologies may be constrained in their ability to be conducted outside of a controlled environment, it does suggest that little of the work being evaluated is at the stage where it can used or distributed in its end user environment.

Moreover, in almost all studies the user learned the system and was tested on its use in the same session. While short studies can provide immediate feedback for further development, they do not account for the time it can take to develop new tactile literacies, become comfortable with a new presentation method, or transcend learning curves that a new technology may have. As such, more longitudinal studies should be considered to truly evaluate new developments in touch-based graphics and to obtain greater buy in and adoption.

\subsubsection{Restriction to functional evaluation:}

The evaluation undertaken in most studies was also strongly focused on function, especially when exploring new presentation methods. Rarely did the studies encapsulate other important factors such as engagement, pleasure or stimulation. Given that an important part of the development of visual graphics is in their aesthetic pleasure, it is suggested that research in this area starts to consider more holistic evaluation criteria.

\subsection{Availability of Outcomes}
As previously stated, commercial work was not considered as part of the literature review. However, in order to gauge the potential ongoing impact of the research in the corpus, the availability of its outcomes was also evaluated during analysis. This was based upon statements in the paper and so probably understates the actual availability of the research.

\begin{table}[h!]
\begin{tabular}{lcc}
\toprule
Availability                      & \multicolumn{2}{c}{Paper Count}    \\
\midrule
Not Available (Custom Prototype)  & 136         &    \\
Available                         & 107         &    \\
\quad Commodity technology              &             & 81 \\
\quad Open source                       &             & 15 \\
\quad Commercialised by the researchers &             & 11 \\
Not applicable / Not reported     & 45          &   \\
 \bottomrule
\end{tabular}
\caption{\label{tab:OutcomeAvailability}Outcome Availability}
\end{table}

Table~\ref{tab:OutcomeAvailability} shows the availability of the outcomes of each research paper in the corpus. Of the work where the nature of availability was clear, almost half the outcomes of the research were considered to be unavailable to the broader research community or end user, due to the research being undertaken with a custom prototype. Close to 40\% of the outcomes are available in some form, whether it be through the use of commodity technologies or with the outcomes being explicitly made available through open source or commercialisation.

%This level of dissemination is quite high. 
It is encouraging that 81 papers did use  commodity technologies. 
%What is important is that the findings from that work are also distributed outside research publication and in ways in which they can have impact for practitioners and end users.
However, it is also acknowledged that there is significant prototyping being undertaken utilising low cost technologies (such as 3D printing, simple electronics, or simple vision systems). This may signal a shift toward research that will not require significant cost to become more widely adopted. As such, continued work should strive to balance innovation with widely adopted commodity technologies as well as with low cost development technologies in order to lead to impact in the BLV community.

\section{Conclusion and Recommendations}
This paper presents a systematic review of literature relating to touch-based graphics over the period of 2010 to mid-2020. Analysis of the work has provided insights into the changing nature of this research, as well as highlighting `calls to action' for researchers in the field. The following are key areas for researchers to consider as well as opportunities for future work:

\begin{itemize}
    \item \emph{Broadening Application Areas:} The areas of application for touch-based graphics should be broadened to include graphics outside the education and O\&M context. While those contexts are fundamental to daily living for the BLV community, it is important for researchers to consider other domain areas. An application area in considerable need of attention is graphics for the workplace. This is not just an opportunity for future research but is also vital for increased support for BLV people in the workplace.
    \item \emph{Strengthening Comparative Studies:} More controlled studies are required in order to better understand the comparative advantages of new presentation methods. Comparative studies can then provide important evidence that can inform decision making regarding adoption of new technologies by the wider BLV community and its stakeholders. These studies should also consider a wider range of graphics.
    \item \emph{Authentic Evaluation:} Laboratory studies dominate the research surveyed. While this is an important first step in research innovation, it is also critical that new developments in touch-based graphics are evaluated in the context in which they will be primarily used and evaluated over longer periods of time. This should be a priority for researchers, as it will also support research becoming more widely adopted in the BLV community.
    \item \emph{BLV Involvement:} While BLV involvement in comparative studies and evaluations is relatively high, there is much less involvement in initial design or prototype evaluation. We suggest that the design of new technologies for touch-based graphics, like other areas of assistive technology design, would greatly benefit from using a more participatory design processes, in particular during the creation of accessible graphics. It is also important that the low vision community is equitably represented in research studies. 
\end{itemize}

It is hoped that this systematic review of work can support the assistive technologies research community by highlighting gaps in current practice as well as opportunities for  work in the future. It forms the basis for a web-based living resource that we hope will be a valuable tool for accessibility researchers for many years to come, and lead to more meaningful impact of this research for the BLV community.

%    \item \emph{Greater Exploration of Tactile Overlays:} Tactile overlays were underrepresented within the corpus. Tactile overlays could provide an excellent way to provide tactility to commodity technologies such as tablets, and as such are a research area that could benefit from greater investigation, especially when created using popular technologies such as 3D printing.

\section{Acknowledgments}
%Anonymised for blind peer review
Cagatay Goncu is supported by the Australian Research Council (ARC) grant DE180100057.

\bibliographystyle{ACM-Reference-Format}
\bibliography{references}


\pagebreak
\clearpage
\onecolumn
\section{Appendix: Comparative Studies}
\begin{longtable}[h!]{|l|p{0.2\textwidth}|p{0.08\textwidth}|p{0.22\textwidth}|p{0.1\textwidth}|p{0.25\textwidth}|}
%\label{tab:comparisons}
\hline
Ref. & Presentation Methods  & Graphic   & Tasks & Participants  & Findings   \\ \hline
\endfirsthead
\multicolumn{6}{c}%
{\tablename\ \thetable\ -- \textit{Continued from previous page}} \\
\hline
Ref. & Presentation Methods  & Graphic   & Tasks & Participants  & Findings   \\ \hline
\endhead
\hline \multicolumn{6}{r}{\textit{Continued on next page}} \\
\endfoot
\hline
\endlastfoot
\cite{Bardot2016}         & Tactile graphic; HT condition: flat surface + hand tracking +  smartwatch feedback (audio + vibration)                   & Map                                                           & Find labelled regions                                                                                                                          & 12 BLV adults                                     & With filtering HT condition faster but without filtering slower; similar correctness; tactile and HT condition with filtering preferred                                                                                                                                           \\ \hline
\cite{Brittell2018}       & Tactile graphic (swell); tactile graphic (embossed); 3D model (printed)                                                 & Map/ Symbol                                                    & Match symbols (point, line, area)                                                                                                                               & 18 BLV adults                                     & More accurate with 3D print than embossed graphic; faster with 3D print than embossed and swell tactile graphic                                                                                                                                                                                                       \\ \hline
\cite{Brock2015}          & Tactile graphic (swell paper); touchscreen + overlay + audio labels            & Map                                                           & Memorise map then answer questions about route, landmark position, layout; tested both short-term and long-term recall                        & 24 BLV adults                                     & Learning time less with audio-labels; slight preference for audio-labels; no effect on short or long-term recall;                                                                                                                                               \\ \hline
\cite{Giraud2017_1}      & Tactile graphic; 3D model (printed) + audio labels                                                                        & Map                                                           & Recall of spatial layout and information in labels                                                                                                            & 24 BLV students                                   & Greater recall of 3D model + audio labels; similar SUS user satisfaction for both                                                                                                                                                                                                                                     \\ \hline
\cite{Giudice2012}        & Tactile graphic; touchscreen + audio + vibration                                                                            & Bar chart; character; geometric shape             & Recall of bar chart layout \& relative bar height; recognise letter; determine shape orientation                                              & 12 sighted (blind-folded) + 3 BLV adults           & Similar accuracy but slight indication tactile more accurate; tactile faster                                                                                                                                                                                                                                              \\ \hline
\cite{Gual-Orti2015}      & Tactile graphic (swell); 3D model (printed)                                                                             & Map/ Symbol                                                    & Flat 2D symbols on tactile graphic vs volumetric 3D symbols on 3D model. Find occurrence of symbol on map.          & 40 BLV + 16 blindfolded sighted adults            & More accurate and faster with 3D volumetric symbol than 2D flat symbol; no difference for 3D printed 2D symbol and 2D symbol on swell paper                                                                                                                                                                           \\ \hline
\cite{Gual2015}           & Tactile graphic (thermoform); 3D model (printed)                                                                        & Map/ Symbol                                                    & Flat 2D symbols on tactile graphic vs volumetric 3D symbols on 3D model. Find occurrence of symbol on map                                                      & 30 BLV + 16 blindfolded sighted adults            & More accurate and faster with 3D volumetric symbol than 2D flat symbol                                                                                                                                                                                                                                                \\ \hline
\cite{Gual2014}           & Tactile graphic (thermoform); 3D model (printed)                                                                        & Map/ Symbol                                                    & Flat 2D symbols on tactile graphic vs mix 2D and volumetric 3D symbols on 3D print. Task: Memorize symbols                                                    & 16 BLV + 6 blindfolded sighted adults             & Better recall of mix of 2D and 3D symbols                                                                                                                                                                                                                                                                             \\ \hline
\cite{Guinness2019}       & Touchscreen + overlay + audio labels; touchscreen + overlay + audio labels + small mobile robots (dynamic shape changing display) & Bar chart                                                     & Identify maximum, minimum and trend.                                                                                                                          & 7 BLV adults                                      & No difference in accuracy or time but preference for the robots                                                                                                                                                                                                                                                       \\ \hline
\cite{Hahn2019}           & Tactile graphic; touchscreen + audio label + vibration                                                                      & Number line, table, pie chart, bar chart, line graph, and map & Education content -- unclear                                                                                                                                  & 22 BLV children + adults (5 excluded)             & No significant difference in accuracy though indication that tactile graphic may be more accurate for bar chart and line graph.                                                                                                                                                                                   \\ \hline
\cite{Hansgen2014}        & Tactile graphic (embossed); tactile graphic (writing film)                                                              & Map                                                           & Readability of braille and other data                                                                                                                          & 21 BLV adults                                     & Overall preference for embossed paper but writing film for durability                                                                                                                                                                                                                                                 \\ \hline
\cite{Holloway2018}                   & Tactile graphic (swell); 3D model (printed)                                                                             & Map                                                           & Understandability of iconic symbols; route finding; memorability; user preferences; understandability                                                          & 16 BLV adults                                     & 3D print preferred and more understandable; 3D iconic symbols easier to understand; better short term recall of 3D print but no difference for longer term recall                                                                                                                                                     \\ \hline
\cite{Kane2013}           & Touchtable; touchtable + overlay (stencils)                                                                               & Map; also input output aids                                   & No fixed tasks-asked to explore                                                                                                                               & 9 BLV adults                                      & Some indication overlay preferred                                                                                                                                                                                                                                                                                     \\ \hline
\cite{Koehler2017}        & tactile graphic; 3D model (print)                                                                                       & Geoscience diagrams /maps explaining tectonic plates             & Build understanding of tectonic plates                                                                                                                        & 5 BLV children                                    & No significant difference in learning outcomes                                                                                                                                                                                                                                                                        \\ \hline
\cite{Lazar2013}          & Touchscreen + overlay + audio labels; touchscreen + audio labels                                                              & Weather Map                                                   & Formative study; no specific tasks                                                                                                                            & 5 BLV adults                                      & Preference for touchscreen + overlay                                                                                                                                                                                                                                                                                    \\ \hline
\cite{Loitsch2012}        & Refreshable tactile display (small) + audio labels; text description                                                      & UML sequence chart                                            & Understanding                                                                                                                                                 & 7 BLV adults                                      & Similar cognitive work load; indication of more errors with textual description                                                                                                                                                                                                                                       \\ \hline
\cite{Melfi2020}          & Tactile graphic + braille index; tactile graphic + digital index; touchscreen + overlay + audio labels                          & Biological diagrams, UML class diagrams                       & Understanding of processes represented by diagrams and recall                                                                                                & 5 BLV adults                                      & Exploration time with tactile graphic + braille index slower than with touchscreen. Results suggest tactile graphic + digital index exploration time lies between these. Results suggest that performance is better with touch screen followed tactile graphics + digital. Indication that students preferred touch screen. \\ \hline
\cite{Papadopoulos2018}   & Verbal description; tablet + overlay + audio labels + environmental audio                                                     & Map                                                           & Navigate in unfamiliar location after orientation training.                                                                                                   & 20 BLV adults                                     & Better performance with tablet                                                                                                                                                                                                                                                                                        \\ \hline
\cite{Ramsamy-Iranah2015} & Tactile graphic (swell); embroidered thread; 3D model (print)                                                           & General/ Symbol                                                & Time to recognise symbol                                                                                                                                      & 27 BLV people age 6-20                            & Recognition time with 3D print less than with tactile graphic, which is less than with embroidered thread.                                                                                                                                                                                                           \\ \hline
\cite{Simmonet2011}       & Tactile graphic (handmade); force feedback + audio labels                                                                 & Map                                                           & Build cognitive map from presentation and then be able to point to landmarks without reference to presentation                                                & 6 BLV adults                                      & Indication that it tool less time to explore tactile graphic; little difference in accuracy.                                                                                                                                                                                                                          \\ \hline
\cite{Toennies2011}       & Touch screen + audio labels;Touch screen + vibration                                                                        & Grid; geometric shapes                                        & Find point on grid ; identify position of point on a grid; identify line and its slope; identify shape and its orientation                                    & 10 sighted people; touch screen hidden from view & No difference on performance. Mediocre performance by both on shape recognition.                                                                                                                                                                                                                        \\ \hline
 \cite{Zeng2015}                        & Tactile graphic (swell); Refreshable braille display + audio labels; touchscreen + audio labels                             & Map                                                           & Build mental map from representation; after viewing presentation build a street map from memory; use representation to plan route; user ranking of difficulty & 10 BLV adults                                     & Longer exploration time with touchscreen; less accurate mental map with touchscreen; less accurate route finding with touchscreen; preferred tactile graphic then refreshable braille display then touchscreen.                                                                                                       \\ \hline
\cite{Zhang2017}          & Tactile graphic (swell); force feedback + audio label + vibration                                                           & Histology image                                               & Differentiate between different types of blood cells                                                                                                          & 5 BLV adults + 5 blindfolded sighted adults       & Blindfolded participants faster with tactile graphic; indication of less accuracy with tactile graphic; BLV participants faster with tactile graphic but less accurate.                                                                                                                                                \\ \hline

\caption{\label{tab:comparisons}Comparative Studies}
\end{longtable}

\end{document}